\documentclass[prc,twocolumn,showpacs,amsmath,amssymb]{revtex4}
\usepackage{rotating}
\usepackage{graphicx}
\usepackage{dcolumn}
\usepackage{amstext}
\usepackage{graphicx}
\usepackage{lscape}
\usepackage{epstopdf}
\usepackage{verbatim}
\usepackage[colorlinks,citecolor=blue,bookmarks]{hyperref}
\usepackage{times}
\usepackage{color}
\usepackage{multirow}
\usepackage[normalem]{ulem}

\begin{document}

\title{Relative mass distributions of neutron-rich thermally fissile nuclei within statistical model} 
 
\author{Bharat Kumar$^{1,4}$}
\email{bharat@iopb.res.in}
\author{M.T. Senthil Kannan$^{2}$}
\author{M. Balasubramaniam$^{2}$}
\author{B. K. Agrawal$^{3,4}$}
\author{S. K. Patra$^{1,4}$}

\affiliation{$^{1}$Institute of Physics, Sachivalaya Marg, Bhubaneswar - 751005, India.}
\affiliation{$^{2}$Department of Physics, Bharathiar University, Coimbatore - 641046, India.}
\affiliation{$^{3}$Saha Institute of Nuclear Physics, 1/AF,  Bidhannagar, Kolkata - 700064, India.}
\affiliation{$^{4}$Homi Bhabha National Institute, Anushakti Nagar, Mumbai - 400094, India.}
\date{\today}

\begin{abstract}
We study the binary mass distribution for the recently predicted thermally 
fissile neutron-rich
uranium and thorium nuclei using statistical model.  The level density parameters needed for the study
 are evaluated from the excitation energies of temperature
dependent relativistic mean field formalism. The excitation energy and the level density
parameter for a given temperature are employed in the convolution integral method to 
obtain the probability of the particular fragmentation. As representative cases, we present the results for the binary yield of $^{250}$U and $^{254}$Th.
The relative yields are  presented for three different temperatures $T =$ 1, 2 and 3 MeV.  
\end{abstract}
\maketitle

\section{Introduction}

Fission phenomenon is one of the most interesting subject in the field of nuclear physics. To study the fission 
properties, a large number of models have been proposed. The fissioning of a nucleus is
successfully explained by the liquid drop model and the semi-empirical mass formula is  the best and simple oldest tool to get a rough estimation of the energy released in a fission process. The pioneering
work of Vautherin and Brink \cite{vat70}, who has applied the Skyrme interaction in a self-consistent method
for the calculation of ground state properties of finite nuclei opened a new dimension in the quantitative estimation of nuclear properties. Subsequently, the Hartree-Fock and time dependent Hartree-Fock 
formalisms \cite{pal} are also implemented to study the properties of fission. Most recently, the microscopic 
relativistic mean field approximation, which is another successful theory in nuclear physics is also
used for the study of nuclear fission \cite{skp10}.

From last few decades, the availability of neutron rich nuclei in various laboratories across the globe opened up new research in the field of nuclear physics, because of their exotic decay properties. The effort
for the synthesis of superheavy nuclei in the laboratories like, Dubna (Russia), 
GSI (Germany), RIKEN (Japan) and BNL (USA) is also quite remarkable. Due to all these, the periodic table is extended
till date upto atomic number $Z = 118$ \cite{ogna}. The decay modes of these superheavy nuclei 
are very different than the usual
modes. Mostly, we understand that, a neutron rich nucleus has a large number of neutron than the
light or medium mass region of the periodic table. The study of these neutron-rich superheavy nuclei is
very interesting, because of their ground state structures and various mode of decays, including
multi-fragment fission (more than two) \cite{skp10}. Another interesting feature of some neutron rich uranium
and thorium nuclei is that similar to $^{233}$U, $^{235}$U and $^{239}$Pu, the nuclei $^{246-264}$U 
and $^{244-262}$Th are also thermally fissile, which are extremely important for the energy production 
in fission process. If the neutron rich uranium and thorium nuclei are the viable sources, then these nuclei
will be more effective to achieve the critical condition in a controlled fission reaction. 

Now the question arises, how we can get a reasonable estimation of the mass yield in the 
spallation reaction of these neutron rich thermally fissile nuclei. As mentioned earlier in this section, there
are many formalisms available in the literature to study these cases. Here, we adopt the statistical
model developed by Fong \cite{fon56}. The calculation is further extended by Rajasekaran and Devanathan \cite{mrd81} to study the binary mass distributions using the single particle energies 
of the Nilsson model. The obtained results are well in agreement with the experimental data.
In the present study, we would like to replace the single particle energies with the excitation
energies of a successful microscopic approach, the relativistic mean field (RMF) formalism. 

For last few decades, the relativistic mean field (RMF) formalism
\cite{wal74,seort86,horo 81,gam90,patra91} with various parameter
sets have successfully reproduced the bulk properties, such
as binding energies, root mean square radii, quadrupole deformation
etc. not only for nuclei near the $\beta-$stability line but also for
nuclei away from it. Further, the RMF formalism is successfully applied
to the study of clusterization of known cluster emitting heavy nucleus
\cite{aru05,bks06,skp07} and the fission of hyper-hyper deformed $^{56}$Ni
\cite{rkg08}. Rutz \textit{et. al.} \cite{rutz95} reproduced the double, triple 
humped fission barrier of $^{240}$Pu, $^{232}$Th and the asymmetric ground 
states of $^{226}$Ra using RMF formalism. Moreover, the symmetric and asymmetric fission
modes are also successfully reproduced. Patra \textit{et. al.} \cite{skp10} studied
the neck configuration in the fission decay of neutron rich U and Th
isotopes. 
The main goal of this present paper is to understand the binary fragmentation 
yield of such neutron rich thermally
fissile superheavy nuclei. $^{250}$U and $^{254}$Th are taken for 
further calculations as the representative cases. 

The paper is organized as follows:   In Section \ref{sec2}, the statistical model
and relativistic mean field theory are presented briefly. In subsection A of this section,
the level density parameter and it's relation with the relative mass yield are outlined. In
subsection B of \ref{sec2}, the equation of motion of the nucleon and meson fields obtained from the
relativistic mean field Lagrangian and the temperature dependent of the equations are adopted through
the occupation number of protons and neutrons. The results are discussed in Section \ref{sec3} and
compared with the finite range droplet model (FRDM) predictions. The summary and concluding remarks 
are given in Section \ref{se4}.  

\section{Formalism} \label{sec2}
The possible binary fragments of the considered nucleus is obtained by equating the charge to mass 
ratio of the parent nucleus to the fission fragments as \cite{mbs2014}:
\begin{equation}
\frac{Z_P}{A_P} \approx \frac{Z_i}{A_i}, \label{eq1}
\end{equation}
with $A_P$, $Z_P$ and $A_i$, $Z_i$ ($i$ = 1 and 2) correspond to mass 
and charge numbers of the parent nucleus and the fission fragments \cite{mrd81}. The 
constraints, $A_1 + A_2 = A$, $Z_1 + Z_2 = Z$ and $ A_1 \ge A_2$ are imposed to satisfy the conservation of charge and mass number in a nuclear fission 
process and to avoid the repetition of fission fragments. 
Another constraint i.e., the binary charge numbers from $ Z_2 \ge $ 26 to $Z_1 \le$ 66 is also taken into consideration from the experimental yield \cite{berg71} to generate the combinations, 
	assuming that the fission fragments lie within these charge range.
\subsection{Statistical theory}
The statistical theory \cite{fon56,cole} assumes that the probability of the particular fragmentation is directly proportional to the folded level density $\rho_{12}$ 
of that fragments with the total excitation energy $E^*$, i.e., $P(A_j,Z_j) \propto \rho_{12}(E^*)$. Where,
\begin{equation}
\rho_{12}(E^*) = \int_{0}^{E^*} \rho_{1}(E^*_{1})\,\rho_{2}(E^* - E_1^*)\, dE^*_1, \label{eq2}
\end{equation} 
and $\rho_i$ is the level density
of two fragments ($i$ = 1, 2). The nuclear level density
\cite{bet37,mor72} is expressed as a function of fragment excitation
energy $E^{\ast}_i$ and the single particle level density parameter
$a_i$ which is given as:
\begin{equation}
\rho_i\left(E^{\ast}_i\right) = \dfrac{1}{12} \left(\dfrac{\pi^{2}}{a_i}\right)^{1/4} E^{\ast (-5/4)}_i\exp\left(2\sqrt{a_iE^{\ast}_i}\right).\label{eq3}
\end{equation}
In Refs. \cite{mbs2014,sk16}, we calculate the excitation energies
of the fragments using the ground state single particle energies of finite range droplet model 
(FRDM) \cite{moller97} at a given temperature $T$ keeping the total number of proton and neutron fixed.
In the present study, we apply the self consistent temperature dependent
relativistic mean field theory to calculate the $E^*$ of
the fragments. The excitation energy is calculated  as,
\begin{equation}
E^{*}_{i}(T) = E_{i}(T) - E_{i}(T = 0). \label{eq4}
\end{equation}
The level density parameter $a_i$ is given as,
\begin{equation}
a_i = \frac{E^*_i}{T^{2}}. \label{eq5}
\end{equation}
The relative yield is calculated as the ratio of the probability
of a given binary fragmentation to the sum of the probabilities of
all the possible binary fragmentations and it is given by,
\begin{equation}
Y(A_j,Z_j)=\frac{P(A_j,Z_j)}{\sum_{j} P(A_j,Z_j)}, \label{eq6}
\end{equation}
where $A_j$ and $Z_j$ are referred to the binary fragmentations involving two fragments with
mass and charge numbers $A_1$, $A_2$ and $Z_1$, $Z_2$ obtained from Eq. \eqref{eq1}.   
The competing basic decay modes such as neutron/proton emission, $\alpha$
decay and ternary fragmentation are not considered. In addition to these approximations, we
have also not included the dynamics of the fission reaction, which are really important to get
a quantitative comparison with the experimental measurements. The
presented results are the prompt disintegration of a parent nucleus
into two fragments (democratic breakup). The resulting excitation energy would
be liberated as prompt particle emission or delayed emission, but such
secondary emissions are also ignored.

\subsection{RMF Formalism}
The RMF theory assume that the nucleons interact with each other
via meson fields. The nucleon - meson interaction is given by the
Lagrangian density \cite{patra91,wal74,seort86,horo 81, bogu77,pric87},

\begin{eqnarray}
{\cal L}&=&\overline{\psi_{i}}\{i\gamma^{\mu}
\partial_{\mu}-M\}\psi_{i}
+{\frac12}\partial^{\mu}\sigma\partial_{\mu}\sigma
-{\frac12}m_{\sigma}^{2}\sigma^{2}\nonumber\\
&& -{\frac13}g_{2}\sigma^{3} -{\frac14}g_{3}\sigma^{4}
-g_{\sigma}\overline{\psi_{i}}\psi_{i}\sigma\nonumber\\
&&-{\frac14}\Omega^{\mu\nu}
\Omega_{\mu\nu}+{\frac12}m_{w}^{2}V^{\mu}V_{\mu}
-g_{w}\overline\psi_{i}
\gamma^{\mu}\psi_{i}
V_{\mu}\nonumber\\
&&-{\frac14}\vec{B}^{\mu\nu}.\vec{B}_{\mu\nu}+{\frac12}m_{\rho}^{2}{\vec
	R^{\mu}} .{\vec{R}_{\mu}}
-g_{\rho}\overline\psi_{i}\gamma^{\mu}\vec{\tau}\psi_{i}.\vec
{R^{\mu}}\nonumber\\
&&-{\frac14}F^{\mu\nu}F_{\mu\nu}-e\overline\psi_{i}
\gamma^{\mu}\frac{\left(1-\tau_{3i}\right)}{2}\psi_{i}A_{\mu}.
\label{eq:7}
\end{eqnarray}
Where, $\psi_i$ is the single particle Dirac spinor. The arrows over the letters in the above equation represent the isovector quantities. The nucleon, the $\sigma$, $\omega$, and $\rho$ meson masses are denoted by M, $m_\sigma$, $m_\omega$ and $m_\rho$ respectively.
The meson and the photon fields are termed as 
$\sigma$, $V_\mu$, $R^\mu$ and $A_\mu$ for $\sigma$, $\omega$, $\rho-$ mesons and photon respectively. 
The $g_{\sigma}$, $g_\omega$, $g_\rho$ and $\frac{e^2}{4\pi}$ are the coupling constants for the $\sigma$, 
$\omega$, $\rho-$mesons and photon fields with nucleons respectively. The strength of the constants 
$g_2$ and $g_3$ are responsible for the nonlinear couplings of $\sigma$  meson ($\sigma^3$ and $\sigma^4$). 
The field tensors of the isovector mesons and the photon are given by,
\begin{eqnarray}
\Omega^{\mu\nu} & = & \partial^{\mu} V^{\nu} - \partial^{\nu} V^{\mu}, \, \\[3mm]
\vec{B}^{\mu\nu} & = & \partial^{\mu} \vec{R}^{\nu} - \partial^{\nu} \vec{R}^{\mu} - g_{\rho} (\vec{R}^{\mu}\times\vec{R}^{\nu}), \,  \\[3mm]
F^{\mu\nu} & = & \partial^{\mu} A^{\nu} - \partial^{\nu} A^{\mu}.	\, 
\end{eqnarray}
The classical variational principle gives the Euler-Lagrange equation and  we get the 
Dirac-equation with potential terms for the nucleons and Klein-Gordan equations with 
source terms for the mesons. We assume the no-sea approximation, so we neglect the 
antiparticle states. We are dealing with the static nucleus, so the time reversal 
symmetry and the conservation of parity simplifies the calculations. After simplifications, 
the Dirac equation for the nucleon is given by,
\begin{equation}
\{ - i\alpha.\bigtriangledown + V(r) + \beta\left[M + S(r)\right]\}\,\,\psi_i = \epsilon_i \, \psi_i, \label{eq11}
\end{equation}
where V(r) represents the vector potential and S(r) is the scalar potential,
\begin{eqnarray}
V(r) &=& g_\omega \omega_0 + g_\rho \tau_{3} \rho_0(r)+ e \frac{(1-\tau_3)}{2} A_0(r), \nonumber \\ [2mm]
S(r) & =& g_\sigma \sigma(r),
\end{eqnarray}
which contributes to  the effective mass,
\begin{equation}
M^*(r) = M + S(r).
\end{equation}

The Klein-Gordon equations for the mesons and the electromagnetic fields with the nucleon densities as sources are,
\begin{center}
	\begin{eqnarray}
	\{-\triangle + m_\sigma^2\}\sigma(r) = -g_{\sigma}\rho_s(r)  -g_2\sigma^2(r)&-\, g_3\sigma^3(r), \label{eq14}\\[2mm] 
	\{-\triangle + m_\omega^2\}\omega_0(r) = g_\omega\rho_v(r), & \\[2mm] \label{eq15}
	\{-\triangle + m_\rho^2\}\rho_0(r) = g_\rho\rho_3(r), &\\[2mm] \label{eq16}
	-\triangle A_0(r) = e\rho_c(r). & \label{eq17}
	\end{eqnarray}	
\end{center}
The corresponding densities such as scalar, baryon (vector), isovector and proton (charge) are given as
\begin{eqnarray}
\rho_s(r) & = &
\sum_i n_i \, \psi_i^\dagger(r) \, \psi_i(r) \,,
\label{eqFN6} \\[1mm]
\rho_v(r) & = &
\sum_i n_i \,\psi_i^\dagger(r) \, \gamma_0 \,\psi_i(r) \,,
\label{eqFN7} \\[1mm]
\rho_3 (r) & = &
\sum_i n_i \, \psi_i^\dagger(r)\, \tau_3\, \psi_i(r) \,,
\label{eqFN8} \\[1mm]
\rho_{\rm c}(r) & = &
\sum_i n_i \,\psi_i^\dagger(r) \,\left (\frac{1 -\tau_3}{2}
\right) \, \psi_i(r) \,.
\label{eqFN9} 
\end{eqnarray}
To solve the Dirac and Klein-Gordan equations, we expand
the Boson fields and the Dirac spinor in an axially deformed
harmonic oscillator basis with $\beta_0$ as the initial deformation
parameter. The nucleon equation along with different meson equations
form a set of coupled equations, which can be solved by iterative
method. The center of mass correction is
calculated with the non-relativistic approximation. The quadrupole deformation parameter
$\beta_2$ is calculated from the resulting quadrupole moments of the
proton and neutron. The total energy is given by \cite{blunden87,reinhard89,gam90},

\begin{eqnarray}
E(T) =& \sum_i \epsilon_i n_i + E_\sigma + E_{\sigma NL} + E_\omega + E_\rho\nonumber
\\[2mm]& + E_C +E_{pair} + E_{c.m.} - AM,
\end{eqnarray}
with
\begin{equation}
E_\sigma  = -\frac{1}{2}g_\sigma \int d^3r \rho_s(r) \sigma(r),
\end{equation}
\vspace{-3mm}
\begin{equation}
E_{\sigma NL} = -\frac{1}{2} \int d^3r \left\lbrace\, \frac{2}{3}g_2 \,\sigma^3(r) + \frac{1}{2}g_3\, \sigma^4(r)\,\right\rbrace,
\end{equation}
\vspace{-3mm}
\begin{eqnarray}
E_\omega = &  -\frac{1}{2}g_\omega \int d^3r \rho_v(r)\omega^0(r),\\[3mm]
E_\rho= &  -\frac{1}{2}g_\rho \int d^3r\rho_3(r)\rho^0(r), \\[3mm]
E_C = &  -\dfrac{e^2}{8\pi}\int d^3r\rho_c(r)A^0(r),
\end{eqnarray}
\vspace{-3mm}
\begin{eqnarray}
E_{pair} = - \triangle\sum_{i>0}u_{i}v_{i} = -\frac{\triangle^2}{G}, 
\end{eqnarray}
\vspace{-3mm}
\begin{equation}
E_{c.m.}= -\frac{3}{4}\times 41A^{-1/3}.
\end{equation}
Here, $\epsilon_i$ is the single particle energy, $n_i$ is the occupation probability and $E_{pair}$ is the pairing energy obtained from the simple BCS formalism.

\subsection{Pairing and temperature dependent RMF formalism}\label{sec:bcs}

The pairing correlation plays a distinct role in open-shell nuclei.
The effect of pairing correlation is markedly seen with increase
in mass number A. Moreover it helps in understanding the deformation
of medium and heavy nuclei. It has a lean effect on
both bulk and single particles properties of lighter mass nuclei
because of the availability of limited pairs near the Fermi surface.
We take the case of T=1 channel of pairing correlation i.e,
pairing between proton- proton and neutron-neutron. In this case, a nucleon of 
quantum states $\vert jm_z\rangle$ pairs with
another nucleons having same $I_z$ value with quantum states $\vert j-m_z
\rangle$,
 since it is the time reversal partner of the other. In both nuclear and atomic
domain the ideology of BCS pairing is the same. The even-odd mass staggering 
of isotopes was the first evidence of its kind for the pairing energy. 
Considering the mean-field formalism, the violation of the particle number is
seen only due to the pairing correlation. We find terms like $\psi^{\dagger} 
\psi$ (density) in the RMF Lagrangian density but we put an embargo on terms
of the form $\psi^{\dagger}\psi^{\dagger}$ or $\psi\psi$ since it violates
the particle number conservation. 
We apply externally the BCS constant pairing gap approximation for our calculation to take 
the pairing correlation into account. The  pairing interaction energy in terms of occupation 
probabilities $v_i^2$ and $u_i^2=1-v_i^2$ is written 
as~\cite{pres82,patra93}:
\begin{equation}
E_{pair}=-G\left[\sum_{i>0}u_{i}v_{i}\right]^2,
\end{equation}
with $G$ is the pairing force constant. 
The variational approach with respect to the occupation number $v_i^2$ gives the BCS equation 
\cite{pres82}:
\begin{equation}
2\epsilon_iu_iv_i-\triangle(u_i^2-v_i^2)=0,
\label{eqn:bcs}
\end{equation}
with the pairing gap $\triangle=G\sum_{i>0}u_{i}v_{i}$. The pairing gap ($\triangle$) of proton and neutron is taken from the empirical formula \cite{va73,gam90}:
\begin{equation}
\triangle = 12 \times A^{-1/2}.
\end{equation}
The temperature introduced in the partial occupancies in the BCS approximation is given by,
\begin{equation}
n_i=v_i^2=\frac{1}{2}\left[1-\frac{\epsilon_i-\lambda}{\tilde{\epsilon_i}}[1-2 f(\tilde{\epsilon_i},T)]\right],
\end{equation}
with 
\begin{eqnarray}
f(\tilde{\epsilon_i},T) = \frac{1}{(1+exp[{\tilde{\epsilon_i}/T}])} & and \nonumber \\[3mm]
\tilde{\epsilon_i} = \sqrt{(\epsilon_i-\lambda)^2+\triangle^2}.&
\end{eqnarray}

The function $f(\tilde{\epsilon_i},T)$ represents the Fermi Dirac
distribution for quasi particle energy $\tilde{\epsilon_i}$. The chemical potential $\lambda_p (\lambda_n)$
for protons (neutrons) is obtained from the constraints of particle number
equations

\begin{eqnarray}
\sum_i n_i^{Z}  = Z, \nonumber \\
\sum_i n_ i^{N} =  N.
\end{eqnarray}
The sum is taken over all proton and neutron states. The entropy is obtained by,
\begin{equation}
S = - \sum_i \left[n_i\, ln(n_i) + (1 - n_i)\, ln (1- n_i)\right].
\end{equation}
The total energy and the gap parameter are obtained by minimizing the free energy,
\begin{equation}
F = E - TS.
\end{equation}
In constant pairing gap calculations, for a particular value of pairing gap $\triangle$
and force constant $G$, the pairing energy $E_{pair}$ diverges, if
it is extended to an infinite configuration space. In fact, in all
realistic calculations with finite range forces, $\triangle$ is not
constant, but decreases with large angular momenta states above the Fermi
surface. Therefore, a pairing window in all the equations are extended
up-to the level $|\epsilon_i-\lambda|\leq 2(41A^{-1/3})$ as a function
of the single particle energy. The factor 2 has been determined so as
to reproduce the pairing correlation energy for neutrons in $^{118}$Sn
using Gogny force \cite{gam90,patra93,dech80}.

\section{Results and discussions} \label{sec3}

In our very recent work \cite{mts16}, we have calculated the ternary mass distributions 
for $^{252}$Cf, $^{242}$Pu and $^{236}$U with the fixed third fragments A$_3 =~^{48}$Ca, $^{20}$O 
and $^{16}$O respectively for the three different temperatures T = 1, 2 and 3 MeV within the TRMF 
formalism. The structure effects of binary fragments are also reported in Ref. \cite{cinthol}. 
In this article, we study the mass distribution of $^{250}$U and $^{254}$Th as a representative cases from the range of neutron-rich thermally fissile
nuclei $^{246-264}$U and
$^{244-262}$Th. Because of the neutron-rich nature of these nuclei, a large number of neutrons emit
during the fission process. These nucleons help to achieve the critical condition much sooner than the
normal fissile nuclei. 

To assure the predictability of the statistical 
model, we also study the binary fragmentation of naturally occurring 
$^{236}$U and $^{232}$Th nuclei. The possible binary fragments are obtained using the Eq. \eqref{eq1}. To calculate the total binding energy at a given temperature, we use the axially symmetric harmonic  oscillator basis expansion N$_F$ and N$_B$ for the Fermion and Boson wave-functions to solve the Dirac Eq. (\ref{eq11}) and the Klein Gordon Eqs.
(\ref{eq14} - \ref{eq17}) iteratively. It is reported \cite{bhar15} that the effect of basis 
space on the calculated binding energy, quadrupole deformation parameter ($\beta_2$) and the 
rms radii of nucleus are  almost equal for the basis set  N$_F =$ N$_B =$ 12 to 20 in the  mass 
region A $\sim$ 200 . Thus, we use the basis space N$_F =$ 12 and N$_B=$ 20 to study the binary 
fragments up to mass number A $\sim$ 182.
The binding energy is obtained by minimizing the free energy, which gives the
most probable quadrupole deformation parameter $\beta_2$ and the proton (neutron)
pairing gaps $\triangle_p$ ($\triangle_n$) for the given temperature. At finite temperature, 
the continuum corrections due to the
excitation of nucleons to be considered. The level
density in the continuum depends on the basis space N$_F$ and N$_B$
\cite{niu09}. It is shown that the continuum corrections need not be included in the calculations 
of level densities up-to the temperature T $\sim$ 3 MeV \cite{bka00,bka98}. 

\begin{figure}[t]
\includegraphics[width=1.0\columnwidth]{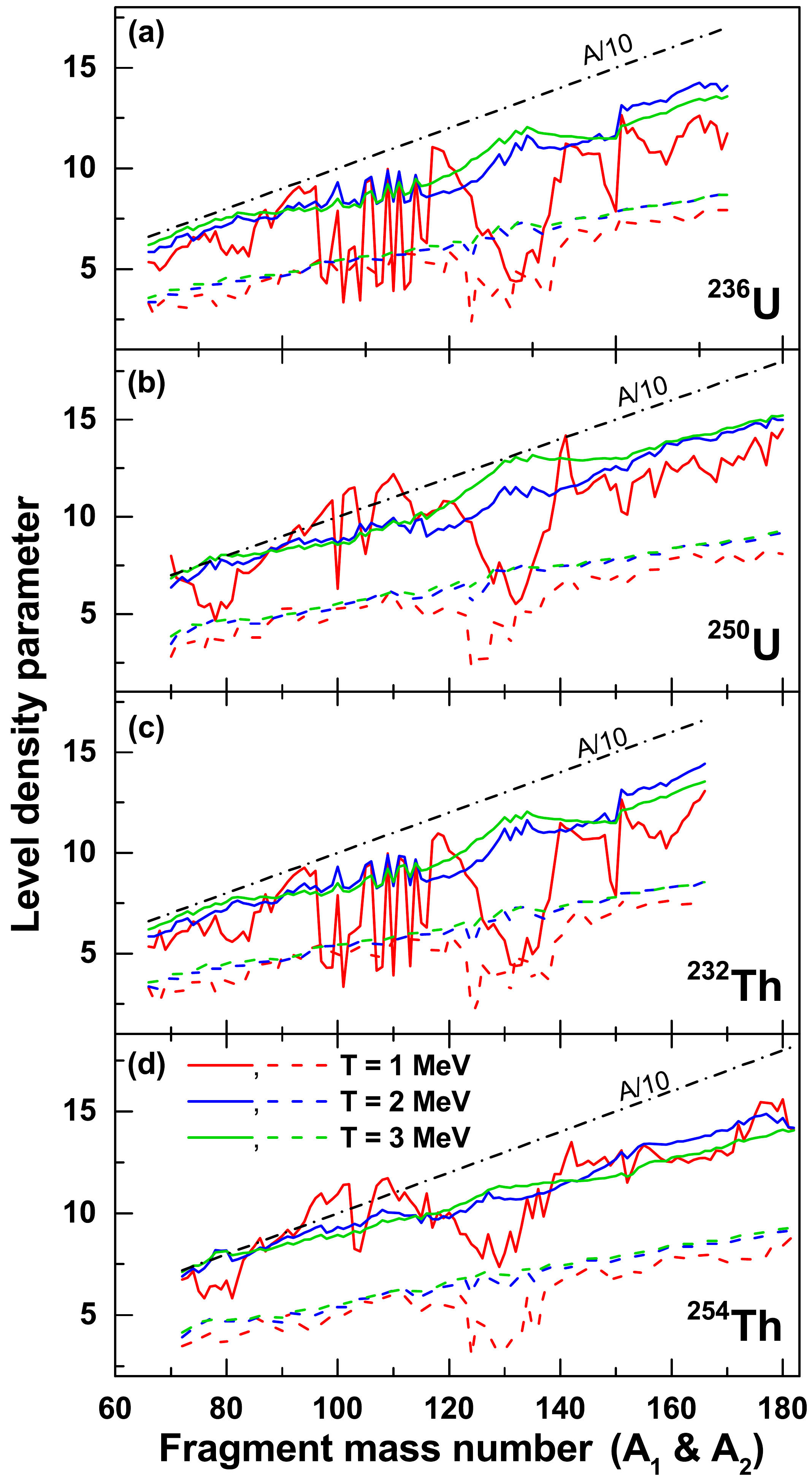}
\caption{(Color online) The level density parameter $a$ for the binary fragmentation of $^{236}$U, $^{250}$U, 
$^{232}$Th and $^{254}$Th at temperature $T =$ 1, 2 and 3 MeV within the TRMF (solid lines) and FRDM (dash lines) formalisms.}
\label{U_ldp}
\end{figure}

\subsection{Level density parameter and level density within TRMF and FRDM formalisms }\label{ldp}

In TRMF, the excitation energies $E^*$ and the level density parameters $a_i$ of the fragments are 
obtained self consistently from Eqns. \eqref{eq4} to \eqref{eq5}. 
The FRDM calculations are  also done for comparison. In this case, level density 
of the fragments are evaluated from the ground state single particle energies of the finite range droplet model 
(FRDM) of M\"{o}ller \textit{et. al.} \cite{moller95} which are retrieved from the Reference Input 
Parameter Library (RIPL-3) \cite{ripl3}. 
The total energy at a given temperature is calculated as $E(T) = \sum n_i \epsilon_i$; $\epsilon_i$ are the 
ground state single particle energies and $n_i$ are the Fermi-Dirac distribution function. The $T$ dependent 
energies are obtained by varying the occupation numbers at a fixed particle number for a given temperature
and given fragment. The level density parameter $a$ is a crucial quantity in the statistical theory for the 
estimation of yields. These values of $a$ for the binary fragments of $^{236}$U, $^{250}$U, $^{232}$Th 
and $^{254}$Th obtained from TRMF and FRDM are depicted in Fig. \ref{U_ldp}. The empirical estimation $a=A/K$ are also
given for comparison,
with $K$, the inverse level density parameter. In general, the $K$ value varies from 8 to 13 with the 
increasing temperature. However, the level density parameter is considered to be  constant up-to $T \approx$ 4 MeV. 
Hence, we take the practical value of $K = 10$  as mentioned in Ref. \cite{ner2002}. 
The $a$ values of TRMF are close to the empirical level density parameter. The FRDM level density parameters
are appreciably lower than the referenced $a$. Further, in both models at $T =$ 1 MeV, there are more 
fluctuations in the level density parameter due to the shell effects of the fragments. At $T =$ 2 and 3 MeV, 
the variations are small. This may be due to the fact that the shell become
degenerate at the higher temperatures. All fragments becomes spherical at temperature $T \approx$ 3 MeV 
as shown in Ref. \cite{cinthol}. 
\begin{figure}[t]
\includegraphics[width=1.0\columnwidth]{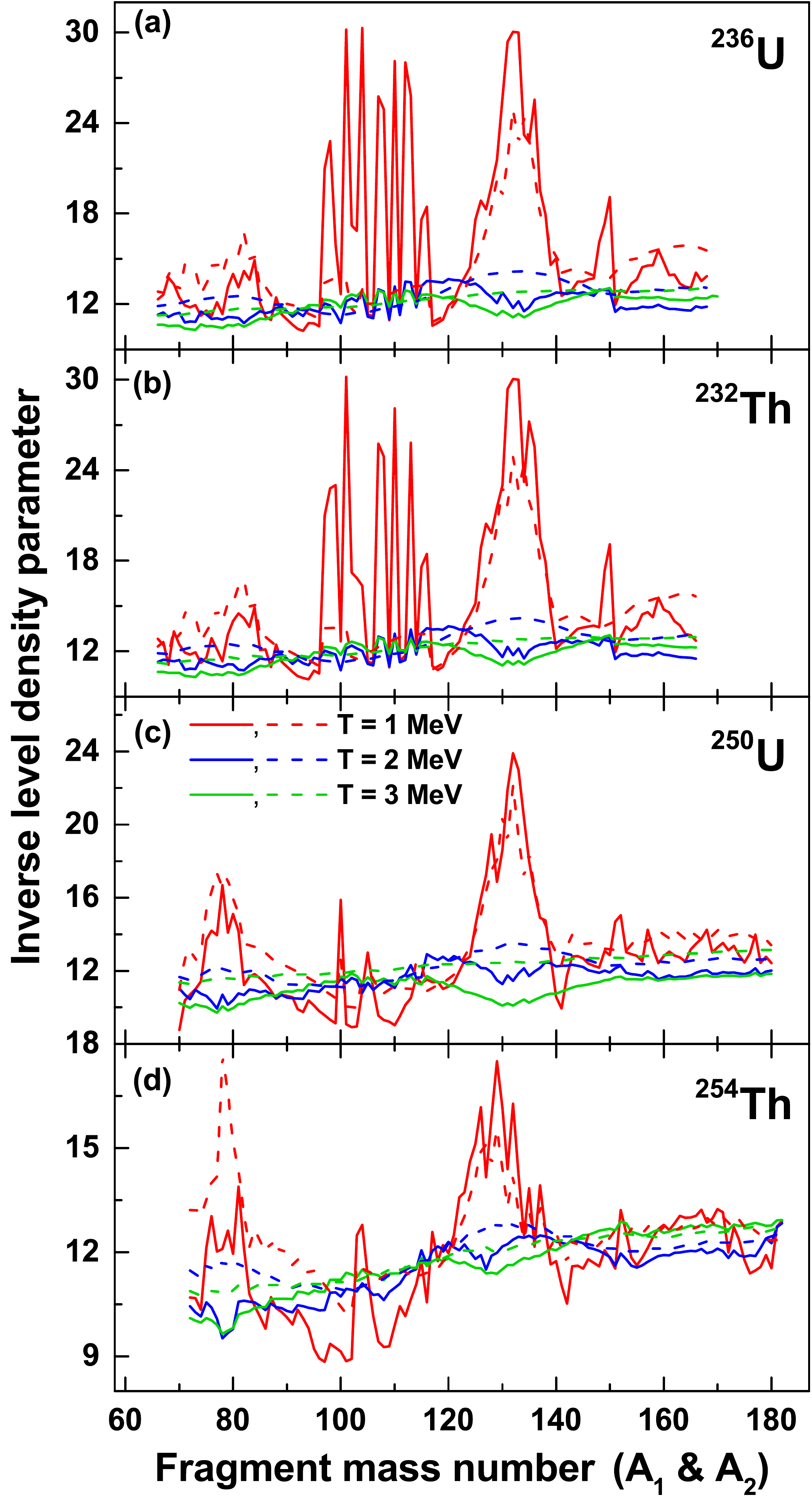}\caption{(Color
online) The inverse level density parameters $K_E$ (solid lines) and $K_S$ (dash lines) are obtained for $^{236}$U, $^{250}$U, $^{232}$Th and $^{254}$Th at temperatures $T =$ 1, 2 and 3 MeV.}
\label{keks}
\end{figure}
\begin{figure}[t]
\includegraphics[width=1.\columnwidth]{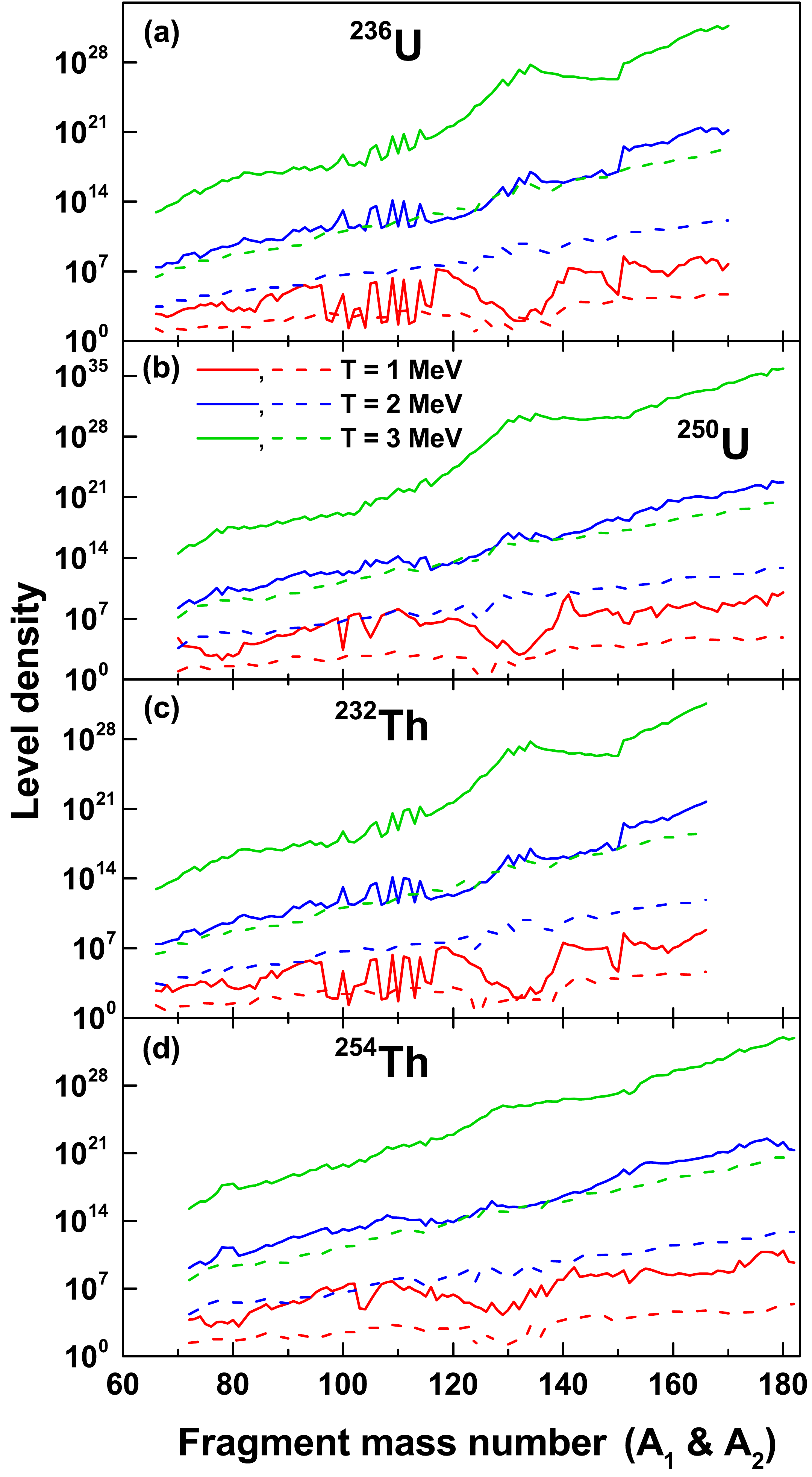}
\caption{(Color online) The level density of the binary fragmentations of $^{236}$U, $^{250}$U, $^{232}$Th 
and $^{254}$Th at temperature $T =$ 1, 2 and 3 MeV  within the TRMF (solid lines) and FRDM (dash lines) formalisms.}
\label{LD}
\end{figure}
The level density parameter $a$ is evaluated in two different ways using excitation energy and the entropy of 
the system as:
\begin{eqnarray}
a_E = \dfrac{E^*}{T^2}, \\ \nonumber
a_S = \dfrac{S}{2T}. 
\end{eqnarray} 
 For instance, the inverse level density parameters $K_E$ and $K_S$ of $^{236}$U, $^{250}$U, $^{232}$Th 
and $^{254}$Th within TRMF formalism are depicted in Fig. \ref{keks}. Both $K_S$ and $K_E$ have maximum
fluctuation upto 30 MeV at $T = $1 MeV. These values reduce to $10-13$ MeV at temperature $T = 2$ MeV or above. It is to be noted 
that at $T =$ 3 MeV, the inverse level density parameter substantially lower around the mass number 
$A \sim$ 130 in all cases. This may be due to the neutron closed shell ($N =$ 82) in the fission fragments 
of $^{236}$U and $^{232}$Th and the neutron-rich nuclei $^{250}$U and $^{254}$Th.
The level density for the fission fragments of $^{236}$U, $^{250}$U, $^{232}$Th and $^{254}$Th are plotted as 
a function of mass number in Fig. \ref{LD} within the TRMF and FRDM formalisms at three different 
temperatures $T =$ 1, 2 and 3 MeV. 

The level density $\rho$ has maximum fluctuations at $T =$ 1 MeV for all 
considered nuclei in TRMF model similar to the level density parameter $a$. The $\rho$ values are 
substantially lower at mass number $A \sim 130$ for all nuclei. In Fig. \ref{LD}, one can notice that the 
level density has small kinks in the mass region $A \sim 71-81$ of $^{236}$U  and $A \sim 77-91$  
of $^{250}$U, comparing with the neighboring nuclei at temperature $T =$ 2 MeV. Consequently, the corresponding 
partner fragments have also higher $\rho$ values. 
A further inspection reveals that the level density of the closed shell nucleus around $A \sim$ 130 has 
higher value than the neighboring nuclei for both $^{236,250}$U, but it has lower yield due to the smaller
level density of the corresponding partners. 
At $T =$ 3 MeV, the level density of the fragments around mass number $A \sim $72 and  130 have 
larger values compared to other fragments of $^{236}$U. On the other hand, the level density 
in the vicinity of neutron number $N = 82$ and proton number $Z = 50$ for the fragments of the neutron-rich 
$^{250}$U nucleus is quite high, because of the close shell of the fragments.
This is evident from the small kink in the level density of $^{130}$Cd ($N =$ 82), $^{132}$In 
($N \sim$ 82) and $^{135}$Sn ($Z =$ 50). 
Again, for $^{232}$Th, the level densities are found to be maximum at around mass number  $A \sim $81 and 100 
for $T =$ 2 MeV. In case of $^{254}$Th, the $\rho$ values are found to be large for the fragments around $A \sim$ 78
and 97 at $T =$ 2 MeV. Their corresponding partners have also similar behavior. For higher temperature 
$T =$ 3 MeV, the higher $\rho$ values of $^{232}$Th fragments are notable around  mass number $A \sim 130$. Similarly, for $^{254}$Th, the fission fragments around $A \sim$ 78 has higher level density at $T =$ 3 MeV. In general, the level density increases towards the neutron closed shell ($N =$ 82) nucleus.

\subsection{Relative fragmentation distribution in binary systems }
\begin{figure}[t]
\includegraphics[width=1.\columnwidth]{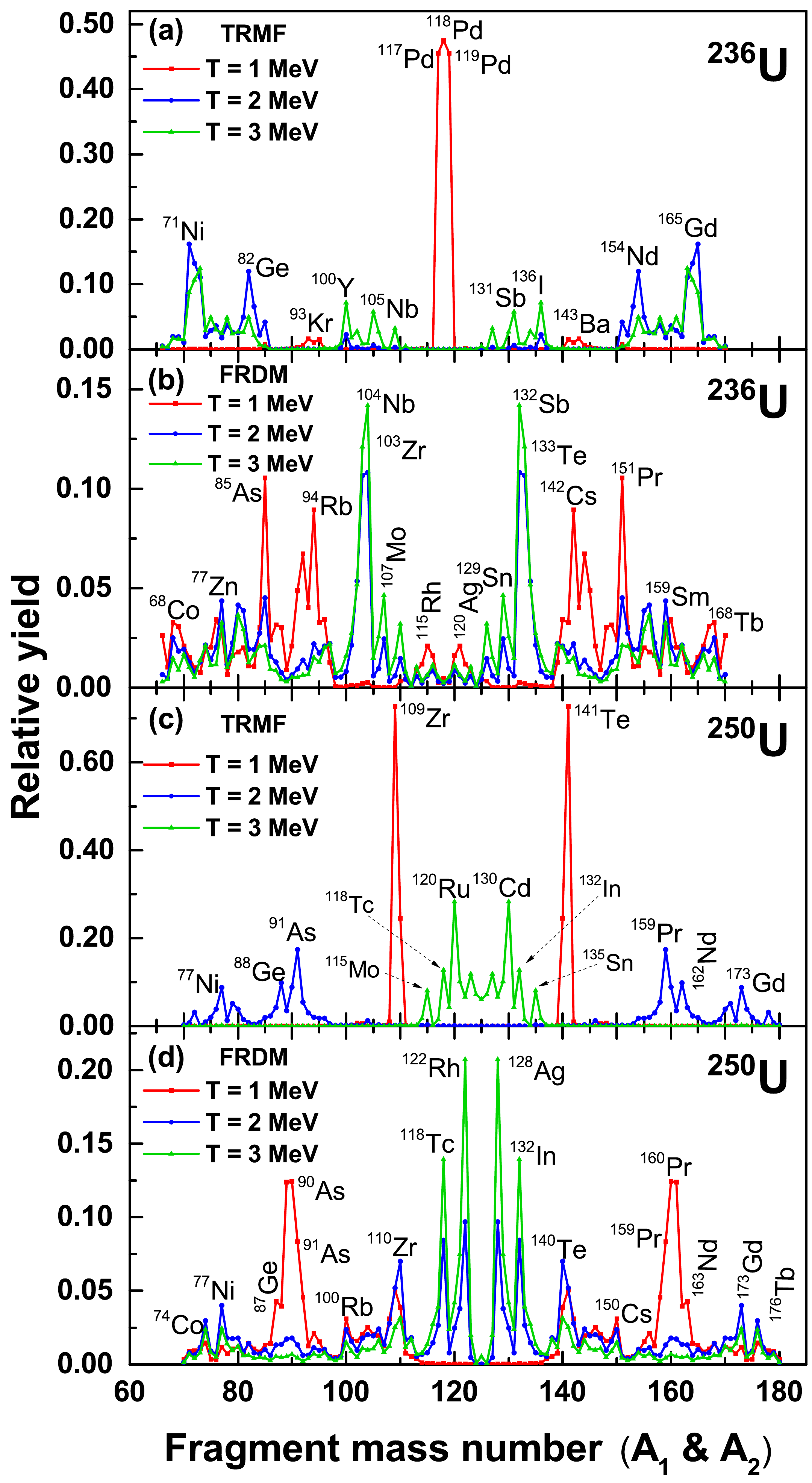} 
\caption{(Color online) Mass distribution of $^{236}$U and $^{250}$U at temperatures T = 1, 2 and 3 MeV. The total yield values are normalized to the scale 2.}
\label{U_yield}
\end{figure}
\begin{figure}[t]
\includegraphics[width=1.\columnwidth]{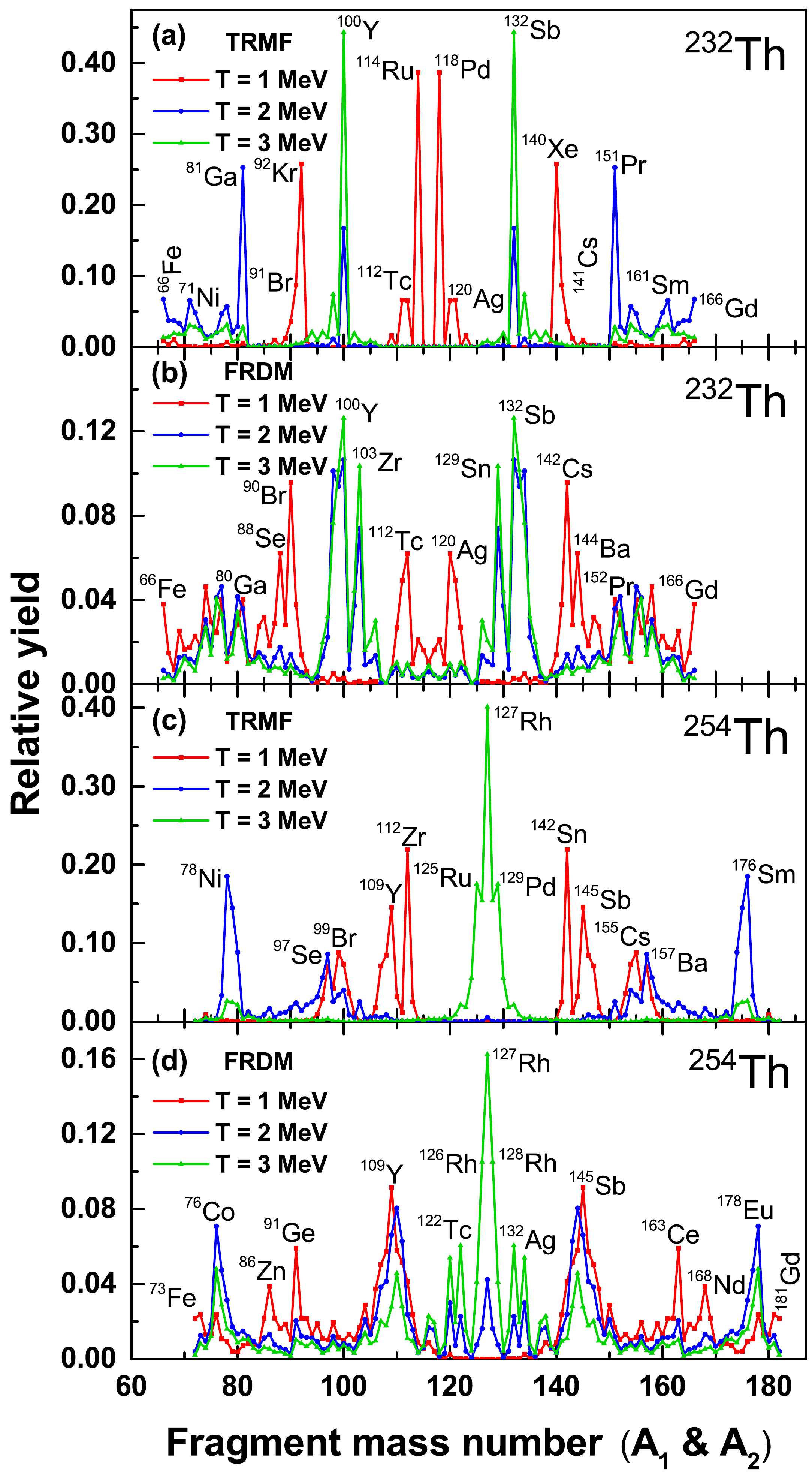}
\caption{(Color online) Mass distribution of $^{232}$Th and $^{254}$Th at temperatures T = 1, 2 and 3 MeV.  The total yield values are normalized to the scale 2.}
\label{Th_yield}
\end{figure} 

In this section, the mass distributions of $^{236}$U, $^{232}$Th and the neutron rich nuclei $^{250}$U and 
$^{254}$Th are calculated at temperatures $T =$ 1, 2 and 3 MeV using TRMF and FRDM excitation energies and 
the level density parameters $a$ as explained in Sec. \ref{sec2}. The binary mass distributions of $^{236,250}$U 
and $^{232,254}$Th are plotted in Figs. \ref{U_yield} and \ref{Th_yield}. The total energy at finite temperature 
and ground state energy are calculated using the TRMF formalism as discussed in the section \ref{ldp}. From the excitation energy E$^{\ast}$ and the temperature $T$, the level density parameter $a$ and the level 
density $\rho$ of the fragments are calculated using Eq. \ref{eq3}. From the fragment level densities $\rho_i$, 
the folding density $\rho_{12}$ is calculated using the convolution integral as in Eq. \ref{eq2} and the relative 
yields are calculated using Eq. \ref{eq6}. The total yields are normalized to the scale 2. 
 
The mass yield of normal nuclei $^{236}$U and $^{232}$Th are briefly explains first, followed by the detailed description of the neutron rich nuclei. The results 
of most favorable fragments yield of $^{236,250}$U and $^{232,254}$Th are listed in Table \ref{tab1} at three different temperatures 
$T =$ 1, 2 and 3 MeV for both TRMF and FRDM formalisms.
From Figs. \ref{U_yield} and \ref{Th_yield}, it is shown that the mass distributions for
$^{236}$U and $^{232}$Th are quite different than the neutron-rich $^{250}$U and $^{254}$Th isotopes. 
 
The symmetric binary fragmentation $^{118}$Pd $+^{118}$Pd for $^{236}$U is the most favorable combination. 
In TRMF, the fragments with close shell ($N =$ 100 and $Z =$ 28) combinations are more probable at the 
temperature $T =$ 2 MeV. The blend region of neutron and proton close shell ($N \approx$ 82 and $Z \approx$ 50) 
has the considerable yield values at $T =$ 3 MeV. The fragmentations $^{151}$Pr $+^{85}$As, $^{142}$Cs $+^{94}$Rb 
and $^{144}$Ba $+^{92}$Kr are the favorable combinations at temperature $T =$ 1 MeV in FRDM formalism. For higher
 temperatures $T =$ 2 and 3 MeV, the closed shell or near closed shell fragments ($N =$  82, 50 and $Z =$ 28) 
have larger yields. From Fig. \ref{Th_yield} in TRMF formalism, the combinations $^{118}$Pd $+^{114}$Ru
 and $^{140}$Xe $+^{92}$Kr are the possible fragments at $T =$ 1 MeV for the nucleus $^{232}$Th. 
At $T =$ 2 MeV, we find maximum yields for the fragments with the close shell or near close shell 
combinations ($N =$ 82, 50).
For higher temperature $T =$ 3 MeV, near the neutron close shell ($N \sim$ 82), $^{132}$Sb $+^{100}$Y is the most 
favorable fragmentation pair compared with all other yields. Similar fragmentations are found in the FRDM formalism 
at $T =$ 2 and 3 MeV. In addition, the probability of the evaluation of $^{129}$Sn $+^{103}$Zr is also quite 
substantial in the fission process. 
For $T =$ 1 MeV, the yield is more or less similar with the TRMF model.

From Fig. \ref{U_yield}, for $^{250}$U the fragment combinations $^{140,141}$Te $+^{110,109}$Zr have the maximum yields at $T =$ 1 MeV in TRMF.
This is also consistent with the evolution of the sub-close shell proton $Z = 40$ in $Zr$ isotopes \cite{blatt}. 
Contrary
to this almost symmetric binary yield, the mass distribution of this nucleus in FRDM formalism have the asymmetric 
evolution of
the fragment combinations like $^{160,159}$Pr $+^{90,91}$As, $^{163,162}$Nd $+^{87,88}$Ge and $^{150}$Cs $+^{100}$Rb. Interestingly, at $T =$ 2 and 3 MeV, the more favorable fragment combinations have one of the closed shell 
nuclei. At $T =$ 2 MeV, $^{159}$Pr $+^{91}$As, $^{162}$Nd $+^{88}$Ge and $^{173}$Gd $+^{77}$Ni are the more probable fragmentations
(see Fig. \ref{U_yield}(c)). It is reported by Satpathy et al \cite{satpathy} and experimentally verified by Patel et al \cite{patel} 
that $N =$ 100 is a neutron close shell for the deformed region, where $Z = 62$ acts like a magic number. 
In FRDM, $^{128}$Ag $+^{122}$Rh, $^{132}$In $+^{118}$Tc, $^{140}$Te $+^{110}$Zr and $^{173}$Gd $+^{77}$Ni have larger yield 
at temperature $T =$ 2 MeV. In TRMF method, the most favorable fragments are confined in the single region 
($A \approx 114-136$) 
which is the blend of vicinity of neutron ($N =$ 82) and proton ($Z =$ 50) closed shell nuclei at $T =$ 3 MeV. 
The fragment combinations $^{130}$Cd $+^{120}$Ru, $^{132}$In $+^{118}$Tc and $^{135}$Sn $+^{115}$Mo are the major
yields for $^{250}$U at T=3 MeV in TRMF calculations.  In FRDM method, at $T =$ 3 MeV, more probable fragments 
are similar that at $T =$ 2 MeV. A comparison between Fig. \ref{U_yield}(c) and 
\ref{U_yield}(d) clears that, although the prediction 
of FRDM and TRMF at $T =$ 3 MeV are qualitatively similar, but it is quantitatively very different at $T =$ 2 MeV 
in both the predictions.
Also, from Fig. \ref{U_yield}, it is inferred that the yields of the fragment combinations in blend region increases and in 
other regions decreases at $T =$ 2 MeV.


In the present study, the total energy of the parent nucleus A is 
more than the sum of the energies of the 
daughters $A_1$ and $A_2$. Here, the dynamics of entire process starting from 
the initial stage upto the scission are ignored. As a result, the energy
conservation in the spallation reaction does not taken into 
account. The fragment yield can be regarded as the relative fragmentation 
probability, which is
obtained from Eq. \ref{eq6}. Now we analyze the fragmentation yields for Th isotopes and the results are depicted in 
Fig. \ref{Th_yield} and Table \ref{tab1}.
In this case, one can see that the mass distribution broadly spreads through out the region $A_i=66-166$. Again,
the most concentrated yields can be divided into two regions I($A_1 =$ 141-148 and $A_2 =$ 106-113) and 
II ($A_1 =$ 152-158 and $A_2 =$ 102-96) for $^{254}$Th in TRMF formalism at the temperature $T =$ 1 MeV. 
The most favorable fragmentation $^{142}$Sn $+^{112}$Zr is obtained from region I. The other combinations in that 
region have also considerable yields. In region II, the isotopes of Ba and Cs appears curiously 
along with their corresponding partners. 
Categorically, in FRDM predictions, region I has larger yields at $T =$ 1 MeV. The other possible 
fragmentations are $^{163}$Ce $+^{91}$Ge, $^{168}$Nd $+^{86}$Zn and $^{181}$Gd $+^{73}$Fe (See Fig. \ref{Th_yield} (b,d)). 
The mass distribution is different with different temperature and the maximum yields at $T =$ 2 MeV in TRMF formalism
are $^{174,175,176}$Sm $+^{80,79,78}$Ni. Apart from these combinations, there are other considerable yields can be
seen in Fig. \ref{Th_yield} for region II.
The prediction of maximum probability of the fragments production in FRDM method are  $^{144}$Sb $+^{110}$Y, 
$^{178}$Eu $+^{76}$Co and $^{127}$Rh $+^{127}$Rh at $T =$ 2 MeV. Besides these yields, one can find other
notable evolution of masses in region I due to the vicinity of the proton close shell. Interestingly, at 
$T =$ 3 MeV, symmetric binary combination $^{127}$Rh $+^{127}$Rh has the largest  yield due to the neutron 
close shell ($N = $ 82) of the fragment $^{127}$Rh.  The other yield fragments have exactly/nearly a magic 
nucleon combination, mostly neutron ($N =$ 82) as one of the fragment. A considerable yield is also seen
for the proton close shell ($Z =$ 28) Ni or/and ($Z=$62) Sm isotopes supporting our earlier prediction \cite{cinthol}. 
This confirms the prediction of Sm as a deformed magic nucleus \cite{satpathy,patel}. 
Another observation of the present calculations show that the yields of the 
neutron-rich nuclei agree with the symmetric
mass distribution of Chaudhuri \textit{et. al.} \cite{cha2015} at large excitation energy, which contradict the recent
prediction of large asymmetric mass distribution of neutron-deficient Th isotopes \cite{pasca}.
These two results \cite{cha2015,pasca} along with our present calculations confirm that the symmetric or asymmetric 
mass distribution at different temperature depends on the proton and neutron combination of the parent nucleus.
In general, both TRMF and FRDM predict maximum yields for both symmetric/asymmetric binary fragmentations followed by 
other secondary fragmentations emission depending on the temperature as well as the mass number of the parent nucleus.
Thus, the binary fragments have larger level density $\rho$ comparing with other nuclei because of neutron/proton 
close shell fragment combinations at $T =$ 2 and 3 MeV. This results ascertain the fact that most favorable 
fragments have larger phase space than the neighboring nuclei as reported earlier \cite{mts16,cinthol}. 
\begin{table*}
	\caption{The relative fragmentation yield (R.Y.) = $Y(A_j,Z_j)=\dfrac{P(A_j,Z_j)}{\sum P(A_j,Z_j)}$ for $^{236}$U, $^{250}$U, $^{232}$Th and $^{254}$Th, obtained with TRMF at the temperatures $T=$ 1, 2 and 3 MeV are 
		compared with the FRDM prediction (The yield values are normalized to 2).}  
	
	\centering
	\renewcommand{\arraystretch}{1.2}
	
	\begin{tabular}{|c|c|cc|cc|c|c|cc|cc|}

		\hline
		
		\multirow{2}{1cm}{Parent} &\multirow{2}{*}{T (MeV)}  & \multicolumn{2}{c|}{TRMF} & \multicolumn{2}{c|}{FRDM} 
		&\multirow{2}{1cm}{Parent} &\multirow{2}{*}{T (MeV)}   & \multicolumn{2}{c|}{TRMF} & \multicolumn{2}{c|}{FRDM } \\\cline{3-6} \cline{9-12}
		
		&&Fragment &R.Y.& Fragment & R.Y.& &&Fragment &R.Y.& Fragment & R.Y.\\
		\hline
		\multirow{11}{*}{$^{236}$U}&\multirow{3}{*}{1}& $^{118}$Pd + $^{118}$Pd  &0.949&  $^{151}$Pr + $^{85}$As  &0.210 & \multirow{11}{*}{$^{250}$U}&\multirow{3}{*}{1}& $^{141}$Te + $^{109}$Zr &1.454&  $^{160}$Pr + $^{90}$As  &0.248 \\
		
		& 	& $^{119}$Pd + $^{117}$Pd  &0.910& $^{142}$Cs + $^{94}$Rb  &0.178&& 	&  $^{140}$Te + $^{110}$Zr &0.491&  $^{161}$Pr + $^{89}$As  &0.247\\
		
		&& $^{143}$Ba + $^{93}$Kr  &0.032& $^{144}$Ba + $^{92}$Kr  &0.134&& 	&  $^{148}$Xe + $^{102}$Sr &0.014&  $^{159}$Pr + $^{91}$As  &0.166\\
		\cline{2-6} \cline{8-12}
		
		&\multirow{3}{*}{2}& $^{165}$Gd + $^{71}$Ni  &0.323& $^{132}$Sb + $^{104}$Nb   &0.216&&\multirow{4}{*}{2}& $^{159}$Pr + $^{91}$As  &0.348& $^{128}$Ag + $^{122}$Rh   &0.193\\	
		
		& &	$^{164}$Gd + $^{72}$Ni  &0.264& $^{133}$Te + $^{103}$Zr  &0.213&	& 	& $^{162}$Nd + $^{88}$Ge  &0.197& $^{132}$In + $^{118}$Tc  &0.168\\	
		
		&&$^{163}$Gd + $^{73}$Ni  &0.0.221 &$^{151}$Pr + $^{85}$As  &0.210&& & $^{160}$Pr + $^{90}$As  &0.176& $^{140}$Te + $^{110}$Zn &0.140 \\
		
		&& $^{154}$Nd + $^{82}$Ge  &0.240& $^{159}$Sb + $^{77}$Zn   &0.087&& & $^{173}$Gd + $^{77}$Ni  &0.175& $^{141}$Te + $^{109}$Zn   &0.100\\
		
		\cline{2-6} \cline{8-12}

		&\multirow{4}{*}{3}& $^{163}$Gd + $^{73}$Ni  &0.249& $^{132}$Sb + $^{104}$Nb   &0.283&&\multirow{4}{*}{3}& $^{130}$Cd + $^{120}$Ru  &0.565& $^{128}$Ag + $^{122}$Rh   &0.414\\	
		
		& 	& $^{164}$Gd + $^{72}$Ni &0.214& $^{133}$Te + $^{103}$Zr  &0.242&	& & $^{132}$In + $^{118}$Tc  &0.255& $^{132}$In + $^{118}$Tc  &0.278\\
		
		&&	$^{136}$I + $^{100}$Y  &0.143& $^{134}$Te + $^{102}$Zr  &0.102&& & $^{127}$Ag + $^{123}$Rh  &0.236& $^{129}$Ag + $^{121}$Rh   &0.149\\
		
		&& $^{131}$Sb + $^{105}$Nb  &0.114& $^{129}$Sn + $^{107}$Mo &0.092& & & $^{135}$Sn + $^{115}$Mo  &0.161& $^{130}$Cd + $^{120}$Ru  &0.083\\\hline	
		
		\hline 
		\multirow{10}{*}{$^{232}$Th}&\multirow{4}{*}{1}& $^{118}$Pd + $^{114}$Ru  &0.773&  $^{142}$Cs + $^{90}$Br  &0.190&\multirow{10}{*}{$^{254}$Th}&\multirow{4}{*}{1}& $^{142}$Sn + $^{112}$Zr  &0.439&  $^{145}$Sb + $^{109}$Y  &0.183\\
		
		& 	& $^{140}$Xe + $^{92}$Kr  &0.515& $^{144}$Ba + $^{88}$Se  &0.124&& 	& $^{145}$Sb + $^{109}$Y  &0.291& $^{163}$Ce + $^{91}$Ge  &0.118\\
		
		&& $^{141}$Cs + $^{91}$Br  &0.174& $^{120}$Ag + $^{112}$Tc  &0.123&& & $^{155}$Cs + $^{99}$Br  &0.176& $^{144}$Sb + $^{110}$Y   &0.115\\
		
		&& $^{120}$Ag + $^{112}$Tc  &0.129& $^{158}$Pm + $^{74}$Cu  &0.092&& 	& $^{157}$Ba + $^{97}$Se  &0.139& $^{168}$Nd + $^{86}$Zn  &0.077\\	
		\cline{2-6} \cline{8-12}
		
		&\multirow{3}{*}{2}& $^{151}$Pr + $^{81}$Ga  &0.505& $^{132}$Sb + $^{100}$Y   &0.213&&\multirow{3}{*}{2}& $^{176}$Sm + $^{78}$Ni  &0.370& $^{144}$Sb + $^{110}$Y   &0.161\\
		
		& &	$^{132}$Sb + $^{100}$Y   &0.334& $^{134}$Te + $^{98}$Sr  &0.202&& 	& $^{175}$Sm + $^{79}$Ni  &0.290& $^{178}$Eu + $^{76}$Co   &0.141\\
		
		&& $^{166}$Gd + $^{66}$Fe  &0.134& $^{129}$Sn + $^{103}$Zr   &0.146&&  & $^{157}$Ba + $^{97}$Se  &0.172& $^{144}$Sb + $^{110}$Y   &0.132\\
		
		\cline{2-6} \cline{8-12}

		&\multirow{3}{*}{3}&  $^{132}$Sb + $^{100}$Y  &0.886&  $^{132}$Sb + $^{100}$Y   &0.252&&\multirow{3}{*}{3}& $^{127}$Rh + $^{127}$Rh  &0.803& $^{127}$Rh + $^{127}$Rh  &0.325\\	
		
		& 	& $^{134}$Te + $^{98}$Sr &0.148& $^{129}$Sn + $^{103}$Zr  &0.207&	& 	& $^{129}$Pd + $^{125}$Ru  &0.350& $^{127}$Rh + $^{127}$Rh  &0.210\\
		
		&& $^{155}$Nd + $^{77}$Zn  &0.063& $^{134}$Te + $^{98}$Sr &0.153&& & $^{128}$Rh + $^{126}$Rh  &0.307& $^{132}$Ag + $^{122}$Tc  &0.120\\	\hline 
		\label{tab1}
	\end{tabular}
\end{table*}

To this end,  it may be mentioned that the differences in the mass
distributions or the relative yields calculated using TRMF and FRDM
approaches  mainly  arise due to the differences in the level densities
associated with these approaches. The mean values and the fluctuations
in the level density parameter   and the corresponding level density are
even qualitatively different in both the approaches considered. This is
possibly stemming from the fact that the single -particle energies  in the
FRDM based model are temperature independent. The temperature dependence
of the excitation energy as required to calculate the level density
parameter  comes only from the modification of the single-particle
occupancy due to the Fermi distribution. In the TRMF approach, the
excitation energy for each fragment at a given temperature is calculated
self-consistently. Therefore, the deformation and the single-particle
energies changes with temperature.

For the neutron-rich
nuclei, the fragments having neutron/proton close shell $N =$ 50, 82 and 100 have maximum possibility of emission
at $T =$ 2 and 3 MeV (for both nuclei $^{250}$U and $^{254}$Th). This is a general trend, we could expect for all 
neutron-rich nuclei. It is worthy to mention some of the recent reports and predictions of multi-fragment
fission for neutron-rich uranium and thorium nuclei. When such a neutron-rich nucleus breaks into nearly two
fragments, the products exceed the drip-line leaving few nucleons (or light nuclei) free. As a result, these free 
particles along with the scission neutrons enhance the chain reaction in a thermonuclear device. These additional 
particles (nucleons or light nuclei) responsible to reach the critical condition much faster than the usual fission 
for normal thermally fissile nucleus. Thus, the neutron-rich thermally fissile nuclei, which are in the case of $^{246-264}$U
and $^{244-262}$Th will be very useful for energy production.

\section{Summary and conclusions} \label{se4}
The fission mass distributions of $\beta-$stable nuclei $^{236}$U and $^{232}$Th and the neutron-rich 
thermally fissile nuclei $^{250}$U and $^{254}$Th are studied within the statistical theory. The possible 
combinations are obtained by equating the charge to mass ratio of the parents to that of the fragments. 
The excitation energies of fragments are evaluated from the temperature dependent self-consistent  
binding energies at the given $T$ and the ground state binding energies which are calculated from the 
relativistic mean field model. The level densities and the yields combinations 
are manipulated from the convolution integral approach. The fission mass distributions of the aforementioned 
nuclei are also evaluated from the FRDM formalism for comparison. The level density parameter $a$ and 
inverse level density parameter $K$ are also studied to see the difference in results with these two
methods. Besides fission fragments, the level densities are also discussed in the present paper. For 
$^{236}$U and $^{232}$Th, the symmetric and nearly symmetric fragmentations are more favorable at temperature 
$T =$ 1 MeV. Interestingly, in most of the cases we find one of the favorable fragment is a close shell
or near close shell configuration ($N =$ 82,50 and $Z =$ 28) at temperature $T =$ 2 and 3 MeV. This result 
ascertains with our earlier predictions. Further, Zr isotopes has larger yield values for $^{250}$U and $^{254}$Th 
with their accompanied possible fragments at $T =$ 1 MeV. The Ba and Cs isotopes with their partners are 
also more possible for $^{254}$Th. This could be due to the deformed close shell in the region $Z=52-66$ 
of the periodic table \cite{jonh}. The Ni isotopes and the neutron close shell ($N \sim$ 100) nuclei are
some of the prominent yields for both $^{250}$U and $^{254}$Th at temperature $T =$ 2 MeV. At $T =$ 3 MeV,
the neutron close shell ($N =$ 82) is one of the largest yield fragments. The symmetric fragmentation 
 $^{127}$Rh $+^{127}$Rh is possible for $^{254}$Th due to the $N = $82 close shell occurs in binary fragmentation. 
For $^{250}$U, the larger yield values are confined to the junction of neutron and proton closed shell nuclei.

\section{Acknowledgment}
The author MTS acknowledge that the financial support from UGC-BSR
research grant award letter no. F.25-1/2014-15(BSR)7-307/2010/(BSR)
dated 05.11.2015 and IOP, Bhubhaneswar for the warm hospitality and for
providing the necessary computer facilities.


\begin{thebibliography}{99}
\bibitem{vat70}
D. Vautherin and D. M. Brink, Phys. Lett. B \textbf{32}, 149 (1970); Phys. Rev. C \textbf{5}, 626 (1972).		
\bibitem{pal} M. K. Pal and A. P. Stamp, Nucl. Phys. A \textbf{99},  228 (1967).
\bibitem{skp10} S. K. Patra, R. K. Choudhury and L. Satpathy, J. Phys. G. 
\textbf{37},  085103 (2010).
\bibitem{ogna} M. Wang, G. Audi, A. H. Wapstra, F. G. Kondev, M. MacCromick,
X. Xu and B. Pfeiffer, Chin. Phys. C \textbf{36}, 1603 (2012).
\bibitem{fon56} P. Fong, Phys. Rev. {\bf 102}, 434 (1956).
\bibitem{mrd81} M. Rajasekaran and V. Devanathan, Phys. Rev. C {\bf 24}, 2606 
(1981).
\bibitem {wal74} J. D. Walecka, Ann. Phys.  {\bf 83}, 491 (1974).
\bibitem {horo 81} C. J. Horowitz and B. D. Serot, Nucl. Phys. A  {\bf 368}, 
503 (1981).
\bibitem{seort86} B. D. Serot and J. D. Walecka, Adv. Nucl. Phys.  {\bf 16}, 1 
(1986).
\bibitem{gam90} Y. K. Gambhir, P. Ring and A. Thimet, Ann. of Phys. {\bf 198}, 
132 (1990).
\bibitem {patra91} S. K. Patra and C. R. Praharaj, Phys. Rev. C  {\bf 44}, 2552 (1991).
\bibitem{aru05} P. Arumugam, B. K. Sharma, S. K. Patra and R. K. Gupta, Phys. 
Rev. C  {\bf 71}, 064308 (2005).
\bibitem{bks06} B. K. Sharma, P. Arumugam, S. K. Patra, P. D. Stevenson, R. K. Gupta and W. Greiner, J. Phys. G.  {\bf 32}, L1 (2006).
\bibitem{skp07} S. K. Patra, Raj. K. Gupta, B. K. Sharma, P. D. Stevenson and W. Greiner, J. Phys. G.  {\bf 34}, 2073 (2007).
\bibitem{rkg08} Raj. K. Gupta, S. K. Patra, P. D. Stevenson, C. Beck, and W. Greiner, J. Phys. G. \textbf{35}, 075106 (2008).
\bibitem{rutz95} K. Rutz, J. A. Maruhn, P.-G. Reinhard and W. Greiner, Nucl. Phys. A \textbf{590}, 680 (1995).
\bibitem{mbs2014} M. Balasubramaniam, C. Karthikraj, N. Arunachalam and S. 
Selvaraj, Phys. Rev. C  {\bf 90}, 054611 (2014); M. Rajasekaran and V. Devanathan, Phys. Rev. C  {\bf 24}, 2606 (1981).
\bibitem{berg71} D. W. Bergen and R. R. Fullwood, Nucl. Phys. A \textbf{163}, 577 (1971).
\bibitem{cole} A. J. Cole, in {\it Fundamental and Applied Nuclear Physics Series - Statistical models for nuclear decay from evaporation to vaporization}, edited by R. R. Betts and W. Greiner, Institute of Physics Publsihing, Bristol and Philadelphia, 2000.
\bibitem{mor72} J. R. Huizenga and L. G. Moretto, Annu. Rev. Nucl. Sci. {\bf 22}, 427 (1972).
\bibitem{bet37} H. Bethe, Rev. Mod. Phys. {\bf 9}, 69 (1937).
\bibitem{sk16} M.T. Senthil Kannan and M. Balasubramaniam, Eur. Phys. J. A  {\bf53}, 164 (2017).
\bibitem{moller97}  P. M\"oller, J. R. Nix, W. D. Myers and W. J. Swiatecki, At. Data and Nucl. Data Tables  {\bf 66}, 131 (1997).
\bibitem{bogu77}  J. Boguta and A. R. Bodmer, Nucl. Phys. A  {\bf 292}, 413 (1977).
\bibitem{pric87} C. E. Price and G. E. Walker, Phys. Rev. C  {\bf 36}, 354 (1987).
\bibitem{blunden87} P. G. Blunden  and  M. J. Iqbal, Phys.  Lett.  B  {\bf 196}, 295 (1987).
\bibitem{reinhard89} P. G. Reinhard,  Rep.  Prog.  Phys. {\bf 52}, 439 (1989). 
\bibitem{patra93} S. K. Patra, Phys. Rev. C  {\bf 48}, 1449 (1993).
\bibitem{pres82} M. A. Preston and R. K. Bhaduri, {\it Structure of Nucleus, Addison-Wesley Publishing Company}, Ch. 8, page 309 (1982).
\bibitem{va73} D. Vautherin, Phys. Rev. C \textbf{7}, 296 (1973).
\bibitem{dech80} J. Decharg\'e and D. Gogny, Phys. Rev. C  {\bf 21}, 1568 
(1980).
\bibitem{mts16} M.T. Senthil kannan, Bharat Kumar, M. Balasubramaniam, B. K. Agrawal, S. K. Patra, Phys. Rev. C \textbf{95}, 064613 (2017).
\bibitem{cinthol} Bharat Kumar, M.T. Senthil kannan, M. Balasubramaniam, B. K. Agrawal and S. K. Patra,  arXiv:1701.00731.
\bibitem{bhar15} Bharat Kumar, S. K. Biswal, S. K.Singh and S. K. Patra, Phys. Rev. C \textbf{92}, 054314 (2015).
\bibitem{niu09} NIU Yi-Fei, LINAG Hao-Zhao and MENG Jie, Chin. Phys. Lett. \textbf{26}, 032103 (2009).
\bibitem{bka00} B. K. Agrawal, Tapas Sil, J. N. De and S. K. Samaddar, Phys. Rev. C \textbf{62}, 044307 (2000); \textit{ibid} \textbf{63}, 024002 (2001).
\bibitem{bka98}  B. K. Agrawal, S. K. Samaddar, J. N. De, and S. Shlomo  Phys. Rev. C \textbf{580}, 3004 (1998).
\bibitem{moller95} P.  M\"oller, J.  R.  Nix  and K. L. Kratz, At. Data and Nucl. Data Tables  {\bf 59}, 185 (1995).
\bibitem{ripl3} https://www-nds.iaea.org/RIPL-3/.
\bibitem{ner2002} B. Nerlo-Pomorska, K. Pomorski, J. Bartel and K. Dietrich, Phys. Rev. C {\bf 66}, 051302(R) (2002).
\bibitem{blatt} J. M. Blatt and V. F. Weisskopf {\it Theoretical Nuclear Physics, Courier Corporation, 1991}.
\bibitem{satpathy}  L. Satpathy and S. K. Patra, J. Phys. G. \textbf{30}, 771 (2004). 
\bibitem{patel} Z. Patel et. al.,  Phys. Rev. Lett. \textbf{113}, 262502 (2014).
\bibitem{cha2015} A. Chaudhuri {\it et. al.}, Phys. Rev. C \textbf{91}, 044620 (2015).
\bibitem{pasca} H. Paşca, A. V. Andreev, G. G. Adamian, and N. V. Antonenko, Phys. Rev. C \textbf{94},  064614 (2016).
\bibitem{jonh} E. F. Jones et. al., J. Phys. G \textbf{30}, L43 (2004).





\end{thebibliography}
\end{document}